\documentclass[10pt,journal,compsoc]{IEEEtran}
\usepackage{comment}
\usepackage{balance}
\usepackage{ragged2e}
\usepackage{booktabs}
\usepackage{tabularx}
\usepackage{orcidlink}
\usepackage{graphicx}
\usepackage{float}
\usepackage{amsmath}
\usepackage{comment}
\usepackage{flushend}
\usepackage{longtable}
\usepackage{tikz}
\usetikzlibrary{positioning, arrows.meta, shapes.geometric}
\usepackage{xcolor}
\usepackage{balance}
\hypersetup{
    colorlinks=true,
    linkcolor=blue,
    citecolor=blue,
    urlcolor=blue
}

\ifCLASSOPTIONcompsoc
  
  \usepackage[nocompress]{cite}
\else
  \usepackage{cite}
\fi
\ifCLASSINFOpdf
\else
\fi

\begin{document}

\title{Mitigating 51\% Attacks in Blockchain Systems Through Early Detection and Checkpoint-Based Defense}

\author{Victor Kebande\,\orcidlink{0000-0003-4071-4596},~\IEEEmembership{Member,~IEEE}
\IEEEcompsocitemizethanks{\IEEEcompsocthanksitem Victor Kebande is with  of University of Colorado, Boulder
CO, 80309, USA. \& University of Colorado Denver, CO, USA\protect\\
E-mail: victor.kebande@ucdenver.edu
}
\thanks{}}

%
%

\markboth{PREPRINT---THIS PAPER HAS BEEN SUBMITTED TO IEEE TRANSACTIONS ON DEPENDABLE AND SECURE COMPUTING}%
{Kebande: Mitigating 51\% Attacks in Blockchain Systems Through Early Detection and Checkpoint-Based Defense}
%



\IEEEtitleabstractindextext{%
\begin{abstract}
The 51\% attack remains one of the  significant security concern in Proof-of-Work (PoW) blockchains, where increasing hash-power concentration can create a malicious  majority-control risk and enable adversarial
chain reorganization. This paper proposes a two-layer defense that combines early hash-power monitoring with checkpoint-based mitigation.
The first layer monitors mining-power concentration and provides an early warning before the critical majority-control threshold is reached.
The second layer uses checkpointing to restrict the depth of accepted chain reorganizations. Monte Carlo simulations are used to
evaluate both mechanisms under different attack scenarios. Across 1,000 simulation runs, a 45\% warning threshold provided a mean warning-to-critical lead time of 10.531 minutes before the modeled 50\% critical threshold was reached. At the selected checkpoint depth of $N=6$, approximately 63.68\% of simulated reorganization attempts
were rejected, while the mean reorganization-depth outcome decreased from 7.948 to 1.634 blocks, representing an approximately 79.4\%
reduction. The results demonstrate that early detection and checkpoint-based mitigation provide complementary mechanisms for reducing the potential impact of 51\% attacks under the evaluated
conditions.

\end{abstract}

}

\maketitle

\IEEEdisplaynontitleabstractindextext

%
\IEEEpeerreviewmaketitle

\IEEEraisesectionheading{\section{Introduction}\label{sec:introduction}}

%
%
%
%
\IEEEPARstart{B}{} lockchain technology enables decentralized systems to establish trust
without relying on a central authority. In Proof-of-Work (PoW)
blockchains. However, security depends heavily on the distribution of
computational power among participating miners \cite{xiao2020modeling}. Excessive concentration
of hash power can expose the network to a 51\% attack, in which an adversary gains sufficient influence to manipulate chain selection,
reverse recent transactions, facilitate double spending, or reorganize previously accepted blocks \cite{saad2020exploring, singh2021blockchain, zheng2023adaptive}. Consequently, hash-power concentration and chain reorganization remain important concerns for the security and finality of PoW in blockchain systems \cite{tao202651}.

A major challenge is that the security problem in blockchain systems does not begin only after an entity reaches the critical majority threshold. Increasing hash-power concentration may provide observable indications of emerging risk before majority control is achieved. Detecting such behavior early could provide
an intervention window. However, detection alone does not prevent an adversary from attempting to reorganize the blockchain. This motivates the need to consider both \emph{early detection} and \emph{attack
mitigation} as complementary components of 51\% attack defense.

This paper investigates whether it is possible  to assess  early hash-power
monitoring  and  checkpoint-based reorganization resistance. The first
layer monitors increasing mining-power concentration and introduces a
warning threshold below the critical 50\% boundary. The second layer uses
checkpoint-based finality to constrain the depth of chain reorganizations.
Together, these mechanisms address two stages of the threat: identifying
dangerous concentration before majority control and limiting the impact
of subsequent reorganization attempts.

Two experiments are conducted to evaluate the approach. The first uses Monte Carlo simulation to investigate warning-to-critical lead time,
detection behavior, and warning-threshold sensitivity under attack,normal, and recovery scenarios coupled with an evaluation under different hash-power growth rates, volatility levels, and attack-progression
patterns. The second evaluates simulated reorganization attempts under checkpoint
depths $N=\{2,4,6,8,10\}$ and a no-checkpoint baseline, examining attack
rejection and accepted reorganization depth. A broader sensitivity analysis
further evaluates checkpoint depths from $N=2$ to $N=13$ under different
attack-depth distributions. Static checkpointing at $N=6$ is also compared
with an adaptive checkpointing policy to characterize the security
trade-offs associated with checkpoint selection.

\subsection{Contributions}
This paper addresses the risks associated with 51\% attacks in
Proof-of-Work blockchain systems through a combination of early detection
and checkpoint-based mitigation. The main contributions of this work are
summarized as follows:
\begin{itemize}
    \item A threat model that characterizes the progression from
hash-power accumulation to majority-control risk and subsequent
chain reorganization.
      \item A two-layer approach combining early hash-power monitoring with
    checkpoint-based mitigation of blockchain reorganizations.
    
    \item A Monte Carlo evaluation of early-warning thresholds, including
    intervention lead time, detection behavior, and threshold sensitivity.

      \item A robustness analysis under varying hash-power growth, volatility,
    and attack-progression patterns.
    
  \item An evaluation of checkpoint depths $N=\{2,4,6,8,10\}$ against a
    no-checkpoint baseline.
    
     \item A checkpoint sensitivity analysis over $N=2$--$13$, 
    attack-depth distributions and static versus adaptive checkpointing.
\end{itemize}

The results demonstrate that these two mechanisms provide complementary
security functions: early monitoring creates an opportunity for
intervention before a critical hash-power condition, while checkpointing
limits the impact of simulated reorganization attempts.

The remainder of this paper is organized as follows.
Section~II reviews the background and related work.
Section~III presents the threat model, and Section~IV describes
the research methodology. Section~V introduces the proposed
two-layer defense. Section~VI presents the experimental setup
and evaluation. Section~VII discusses the experimental results,
robustness, parameter trade-offs, and limitations. Finally,
Section~VIII concludes the paper and outlines directions for
future work.



\section{Background and Related Work}

In this section, the author explores the background on the topic of mitigating the risks of 51\% attacks in blockchain systems. The primary objective of this background study is to identify and understand the challenges associated with these attacks, as well as the existing strategies for preventing them. To achieve this objective, the study will focus on several key areas, including an overview of the blockchain technology, an explanation of the 51\%  attack, and a review of the existing prevention strategies. 

\begin{table}[!h]
\centering
\caption{Summary of Vulnerabilities Associated with 51\% Attacks}
\label{tab:vulnerabilities}

\scriptsize
\setlength{\tabcolsep}{3pt}
\renewcommand{\arraystretch}{1.0}

\begin{tabularx}{\columnwidth}{c p{2.2cm} X}
\hline
\textbf{No.} &
\textbf{Vulnerability} &
\textbf{Description} \\
\hline

1 & Double Spending \cite{nicolas2019novel,mukherjee2025double} &
Spending the same asset twice through manipulation of blockchain history. \\

2 & Transaction Reversal \cite{mukherjee2025double} &
Reversing previously confirmed transactions through chain reorganization. \\

3 & Ledger Reorganization \cite{sello2025rethink,mukherjee2025double} &
Replacing part of the accepted blockchain history with an alternative chain. \\

4 & Blocking of Transactions \cite{aponte202151,sello202651} &
Preventing or delaying legitimate transactions from being confirmed. \\

5 & Transaction Censorship \cite{sello202651} &
Selectively excluding transactions from blocks or preventing their confirmation. \\

\hline
\end{tabularx}
\end{table}

\subsection{Proof-of-Work Blockchain Security}
\label{subsec:pow-security}

Proof-of-Work (PoW) blockchains achieve distributed consensus by requiring
miners to expend computational resources when competing to append new blocks
to the blockchain \cite{shi2022pooling}. The probability that a miner successfully extends the
chain is therefore closely related to the proportion of the total network
hash power under its control. This mechanism enables mutually untrusted
participants to agree on a common transaction history without relying on a
central authority \cite{xu2024exploring}.

The security of PoW, however, depends not only on the cryptographic
properties of the underlying hash function but also on the distribution of
computational power among participating miners \cite{wu2026defending}. When mining power is
sufficiently decentralized, no individual miner or coordinated group can
easily dominate block production. As hash power becomes increasingly
concentrated, this assumption weakens and the probability that a dominant
entity can influence chain selection increases \cite{khan2026consensus}.

PoW blockchains generally resolve competing branches by following the chain
with the greatest accumulated proof of work. Consequently, an adversary with
substantial computational resources may attempt to construct an alternative
branch and cause previously accepted blocks to be reorganized. Such a
reorganization can affect transaction finality and, under appropriate attack
conditions, facilitate transaction reversal or double spending \cite{zheng2023adaptive, mukherjee2025double}.
The security concern is therefore not limited to whether an adversary has
already obtained majority computational power; the progression toward such
concentration is itself relevant to defensive monitoring.

\subsection{Hash-Power Concentration and 51\% Attacks}
\label{subsec:51-attacks}

A 51\% attack refers to a majority-control scenario in which an adversary,
or a coordinated group of miners, obtains sufficient computational power to
exert dominant influence over block production and chain selection in a PoW
blockchain \cite{aponte202151, badertscher2021rational, lee2025inside}. Although the attack is conventionally associated with
the 50\% threshold, reaching this value should not be interpreted as an
automatic guarantee of a successful attack. Attack outcomes also depend on
factors such as network conditions, competing miners, attack duration, and
the adversary's ability to sustain its computational advantage.

Let $h_i(t)$ denote the hash rate controlled by miner or mining pool $i$ at
time $t$, and let $H(t)$ denote the total network hash rate. The corresponding
hash-power share can be represented as

\begin{equation}
s_i(t)=\frac{h_i(t)}{H(t)}.
\end{equation}

As $s_i(t)$ increases, the concentration of computational power becomes an
important security indicator. In the threat model considered in this work,
the attack process is therefore treated as progressive rather than
instantaneous. An adversary may accumulate hash power over time, moving the
system from a comparatively decentralized state toward a region of
majority-control risk.

Rather than treating majority control as an instantaneous event, this study
considers progressive hash-power concentration. A 45\% threshold is used as
an early-warning point and 50\% as the critical majority-control threshold.
The interval between these thresholds provides an intervention window that
is evaluated experimentally.

Majority computational influence can also create conditions for blockchain
reorganization. An adversary may privately construct a competing branch and
attempt to replace part of the publicly accepted chain when its alternative
branch accumulates sufficient proof of work \cite{deirmentzoglou2019survey}. The security impact then depends
not only on whether a reorganization occurs but also on its depth. Deeper
reorganizations can invalidate a larger portion of previously accepted chain
history and therefore present a greater threat to transaction finality.

\subsection{Security Impact of Majority Control}
\label{subsec:majority-control}

Majority computational control in a PoW blockchain can significantly weaken
the security guarantees provided by decentralized consensus. An adversary
with sufficient hash power can influence chain selection and block production,
creating opportunities for double spending, transaction reversal, transaction
censorship, and chain reorganization \cite{sello202651, lovejoy2020empirical, deirmentzoglou2019survey}. These actions can undermine transaction
finality and reduce confidence in the integrity of the blockchain.

The severity of majority control is particularly evident in chain
reorganization attacks \cite{zhang2024breaking, sello2025rethink}, where an adversary attempts to replace previously
accepted blocks with an alternative chain. The impact of such an attack
depends partly on the depth of the reorganization, since deeper reorganizations
can affect a larger portion of the confirmed transaction history.

These security consequences motivate defenses at different stages of the attack process. Early monitoring of hash-power concentration can provide
warning of increasing majority-control risk, while checkpointing can restrict the depth of chain reorganizations that may be accepted. This combination
forms the basis of the detection and mitigation mechanisms investigated in this work.

\subsubsection{Vulnerabilities in 51\% Attacks}

The 51\% attack is considered a significant vulnerability in blockchain systems and has been observed in various blockchain platforms such as Ethereum Classic, Bitcoin Gold, and Monacoin. The risk of such an attack increases as the blockchain network grows, and more computational power is required to maintain its security.

Vulnerabilities in 51\% attacks are rooted in the decentralized nature of blockchain systems, which is one of the key features that provides security and transparency \cite{sello202651, aponte202151, ye2018analysis}. However, the vulnerability of 51\% attacks is a major concern for the implementation of blockchain technology in critical infrastructure, such as financial systems. Mitigating these risks help to ensure that blockchain systems are reliable and   secure. To achieve this, it is essential to understand the underlying causes and potential consequences of 51\%attacks and to explore potential mitigation strategies that can be employed to secure blockchain systems against these attacks.This situation allows the attacker to control the majority of the network, enabling them to manipulate the system by double-spending, halting transactions, and reversing previously confirmed transactions \cite{zheng2023adaptive, xiao2020modeling}. In other words, a 51\% attack is a situation where the integrity of the blockchain system is compromised, and the consensus mechanism is broken.

These vulnerabilities demonstrate the significant risks that 51\% attacks pose to the security and stability of blockchain systems.

\subsubsection{Existing Detection and Mitigation Approaches}
\label{subsec:existing-approaches}

Existing approaches for addressing 51\% attacks generally focus on either
detecting abnormal mining behavior or limiting the consequences of majority
control. Detection-based approaches monitor indicators such as hash-power
distribution \cite{monrat2019survey, tschorsch2016bitcoin}, mining-pool concentration \cite{tran2024routing}, block-generation patterns \cite{wang2019survey}, and
changes in network behavior to identify conditions associated with increasing
centralization or potential attacks \cite{mosakheil2018security, tran2024routing}. Such mechanisms can provide
early indications of risk, but detection alone does not prevent an adversary
from attempting to reorganize the blockchain.

\begin{figure*}[!t]
    \centering
    \includegraphics[width=\textwidth]{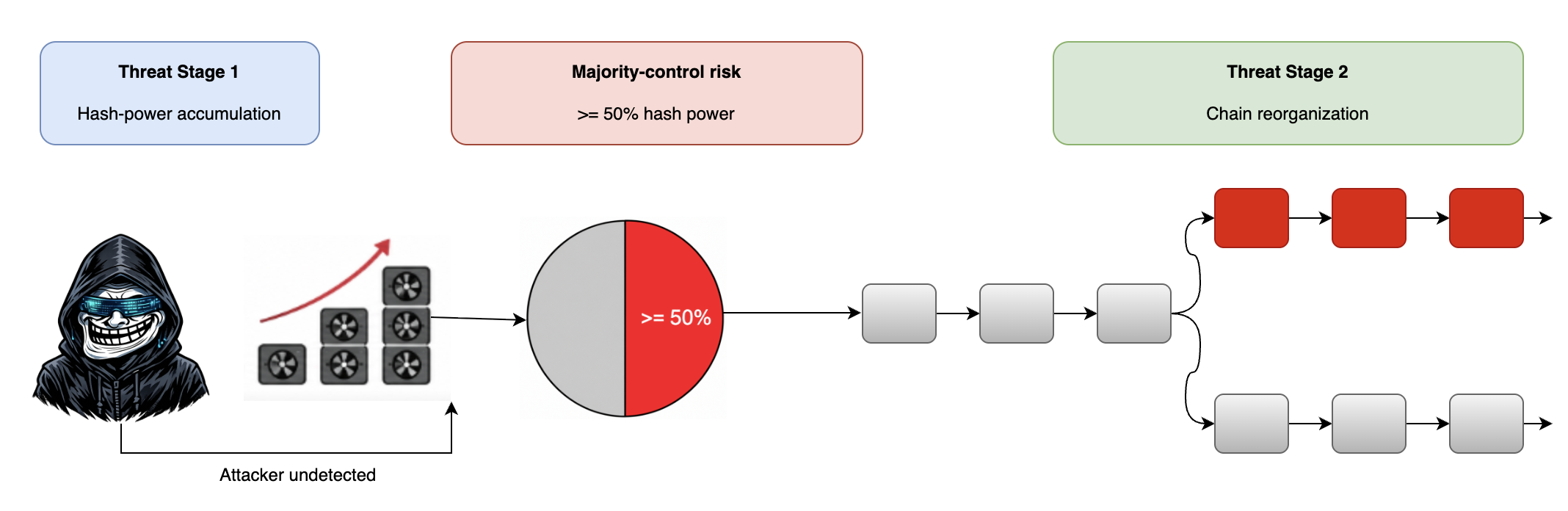}
    \caption{Threat model illustrating hash-power accumulation,
    majority-control risk, and subsequent chain reorganization in a
    51\% attack.}
    \label{fig:threat-model}
\end{figure*}

Mitigation approaches seek to reduce the impact of an attack after
majority-control risk has emerged. These include confirmation requirements,
checkpointing, and finality mechanisms that restrict modifications to
previously accepted blockchain history \cite{sello202651, mukherjee2025double, tao202651}. Checkpointing is particularly
useful for limiting deep chain reorganizations by establishing blocks beyond which alternative chains are not accepted.

Although detection and mitigation have individually received considerable
attention, relying on either mechanism alone provides only partial protection.
Early detection can provide an intervention window but does not directly
constrain chain reorganization, whereas checkpointing can limit
reorganizations but does not provide advance warning of increasing hash-power
concentration. 


\section{Threat Model}
\label{subsec:threat-model}

This study considers a Proof-of-Work (PoW) blockchain in which the
computational power required for consensus is distributed among multiple
miners or mining pools. The adversary seeks to progressively accumulate
a substantial share of the total network hash power with the objective of
gaining majority influence over the blockchain. As illustrated in
Fig.~\ref{fig:threat-model}, the considered threat is divided into two
main stages: hash-power accumulation and chain reorganization.

 \textbf{Threat Stage 1:} The adversary progressively increases its
share of the network hash power. During this stage, the adversary may
remain undetected while its computational influence continues to grow. As the controlled hash power approaches the majority threshold, the risk
to the consensus mechanism increases. Once the adversary controls approximately 50\% or more of the network hash power, it enters the majority-control region considered in this study. This stage motivates
the early-warning mechanism evaluated in Experiment~1, where increasing hash-power concentration is monitored before reaching the critical threshold.

 \textbf{Threat Stage 2:} The adversary uses its majority influence to attempt a chain reorganization. The adversary can construct a competing
chain that conflicts with previously accepted blockchain history and attempt to cause the network to adopt the alternative chain. Successful reorganization may undermine transaction finality and create conditions
for attacks such as transaction reversal and double spending. This stage motivates Experiment~2, in which checkpoint-based finality is evaluated
as a mechanism for restricting the depth of accepted reorganizations.

The adversary is assumed to control mining resources and initiate
reorganization attempts, but is not assumed to compromise cryptographic
primitives, directly modify honest nodes, or alter the defensive
checkpoint mechanism. Other blockchain attacks, including private-key
compromise, eclipse attacks, denial-of-service attacks, and software
implementation vulnerabilities, are outside the scope of this threat
model. Accordingly, the security objectives are twofold: (i) to identify
dangerous hash-power concentration before the critical majority condition
is reached, and (ii) to limit the impact of subsequent chain
reorganizations through checkpoint-based mitigation.

\section{Research Methodology}

\begin{figure*}[!t]
    \centering
    \includegraphics[width=\textwidth]{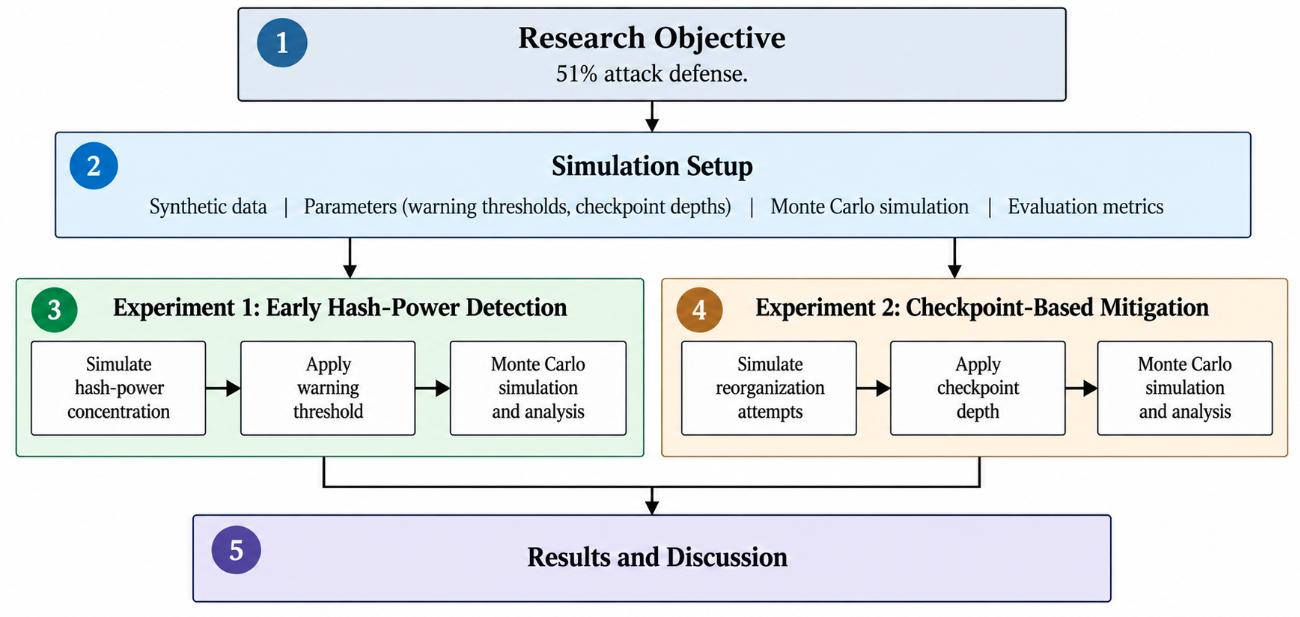}
    \caption{Research methodology for the simulation-based evaluation of early hash-power detection and checkpoint-based mitigation of 51\% attacks.}
    \label{fig:methodology}
\end{figure*}

In this study, a simulation-based quantitative research methodology is
used to investigate the detection and mitigation of 51\% attacks in
blockchain systems. As illustrated in Fig.~\ref{fig:methodology}, the
methodology consists of five main processes: research objective,
simulation setup, early hash-power detection, checkpoint-based mitigation,
and results analysis. The simulation environment uses synthetic data to
model hash-power concentration and chain-reorganization attempts, with
warning thresholds, checkpoint depths, Monte Carlo runs, and evaluation
metrics defined as the main experimental parameters.

The first experiment investigates early detection by simulating increasing
hash-power concentration and applying warning thresholds below the critical
50\% level. Monte Carlo simulations are used to evaluate warning-to-critical
lead time, detection behavior, false warnings, and threshold sensitivity
under attack, normal, and recovery scenarios. The second experiment
investigates checkpoint-based mitigation by generating simulated
reorganization attempts and evaluating them under checkpoint depths
$N=\{2,4,6,8,10\}$ and a no-checkpoint baseline. Attack rejection rates
and accepted reorganization depths are measured across repeated simulations.
Finally, the results from both experiments are statistically analyzed and
compared to determine the effectiveness and parameter trade-offs of early
hash-power detection and checkpoint-based mitigation for 51\% attack
defense.

\section{Proposed Approach}
\label{sec:proposed-approach}

\subsection{Overview}
\label{subsec:approach-overview}

\begin{figure*}[!t]
    \centering
    \includegraphics[width=\textwidth]{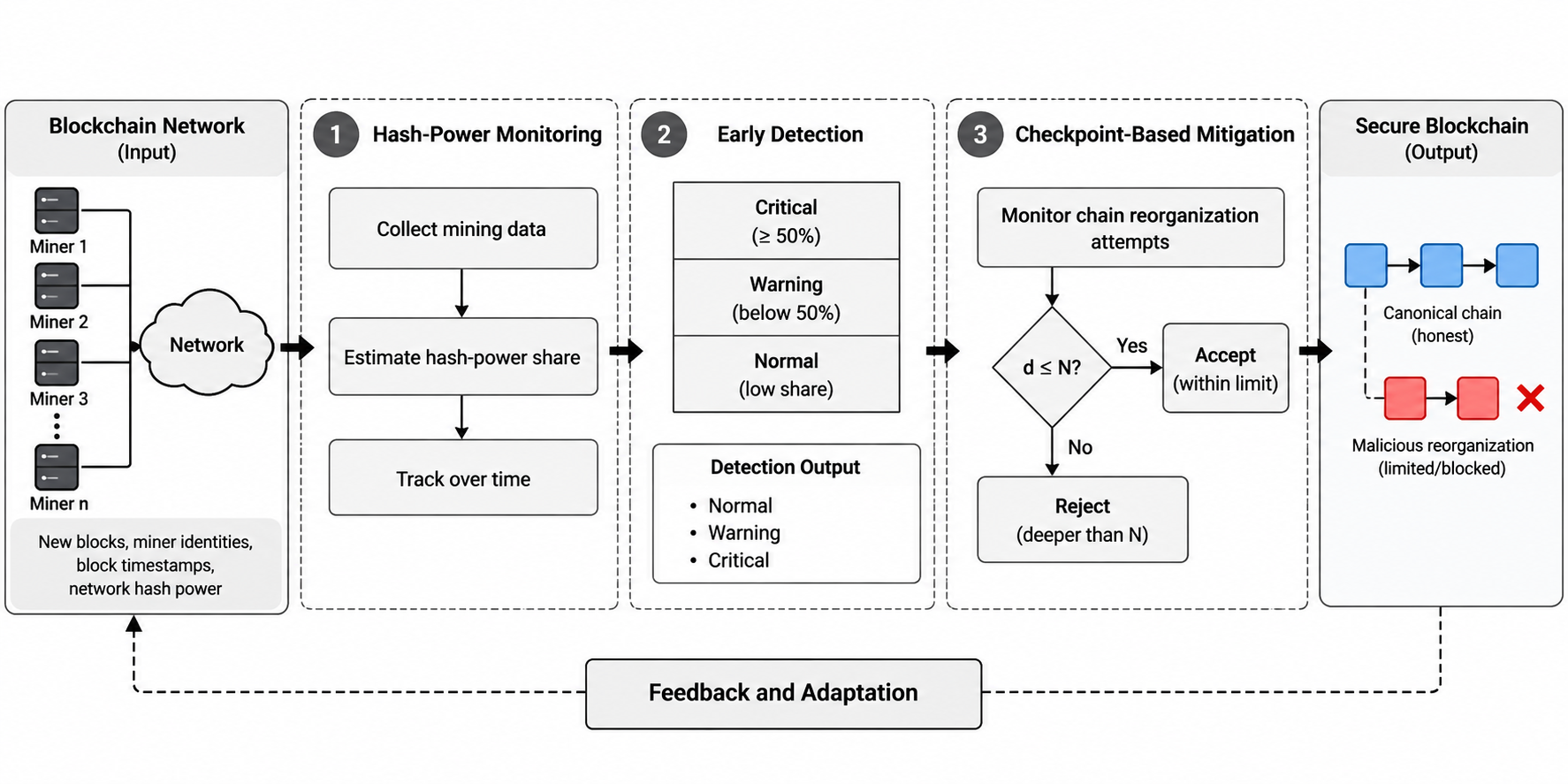}
    \caption{Proposed integrated framework for early hash-power detection and checkpoint-based mitigation against 51\% attacks.}
    \label{fig:proposed-framework}
\end{figure*}

The proposed approach provides a two-layer defense against 51\% attacks by
combining early hash-power detection with checkpoint-based mitigation.
The first layer targets the period in which an adversary progressively
accumulates computational power, while the second layer limits the impact
of a chain-reorganization attempt after the network enters a
majority-control risk state. The approach therefore addresses both the
development of the threat and its potential consequences. As illustrated in Fig.~\ref{fig:proposed-framework}, the proposed framework combines hash-power monitoring, early detection, checkpoint-based mitigation, and feedback-based adaptation.

Let $H_A(t)$ denote the hash power controlled by an adversary at time $t$,
and let $H_T(t)$ denote the total network hash power. The adversary's
relative hash-power share is defined as

\begin{equation}
    S_A(t) = \frac{H_A(t)}{H_T(t)}.
    \label{eq:hash-share}
\end{equation}

The approach continuously observes $S_A(t)$ and compares it with an
early-warning threshold $\theta_w$ and a critical threshold
$\theta_c$. In this study, $\theta_c=0.50$, while
$\theta_w < \theta_c$. If the monitored share approaches the critical
region, an early warning is generated. If majority-control risk develops
and a chain reorganization is subsequently attempted, a checkpoint policy
is applied to restrict the depth of blockchain history that can be
reorganized.

Conceptually, the proposed defense can be represented as

\[
\begin{gathered}
\text{Hash-Power Monitoring} \rightarrow \text{Early Warning} \\
\downarrow \\
\text{Majority-Control Risk} \rightarrow \text{Checkpoint Mitigation}
\end{gathered}
\]

\subsection{Early Hash-Power Detection}
\label{subsec:early-hash-detection}

The first component of the proposed approach is designed to identify
potentially dangerous hash-power concentration before the critical
majority condition is reached. Instead of using the 50\% boundary as the
first indication of risk, an early-warning threshold $\theta_w$ is
introduced below the critical threshold.

For a monitored mining entity or coordinated adversary, the detection
state at time $t$ is defined as

\begin{equation}
D(t)=
\begin{cases}
\text{Normal}, & S_A(t)<\theta_w,\\
\text{Warning}, & \theta_w \leq S_A(t)<\theta_c,\\
\text{Critical}, & S_A(t)\geq\theta_c.
\end{cases}
\label{eq:detection-state}
\end{equation}

The warning state indicates increasing concentration rather than a
successful 51\% attack. Its purpose is to provide an observable interval
during which the network or its operators can recognize the developing
risk before the critical condition is reached. If $t_w$ denotes the first
time at which the warning threshold is crossed and $t_c$ denotes the first
critical-threshold crossing, the available intervention lead time is

\begin{equation}
    T_{\mathrm{lead}} = t_c - t_w.
    \label{eq:lead-times}
\end{equation}

A larger $T_{\mathrm{lead}}$ provides earlier awareness of the developing
risk. However, selecting a very low value of $\theta_w$ may increase the
likelihood of warnings during temporary or non-malicious fluctuations in
hash-power distribution. The warning threshold must therefore balance
early detection against unnecessary warning events.

\subsection{Checkpoint-Based Mitigation}
\label{subsec:checkpoint-mitigation}

Early detection alone cannot prevent an adversary with sufficient
computational influence from attempting to reorganize the blockchain.
The second component therefore introduces checkpoint-based mitigation to
limit the extent of chain history that can be replaced.

Let $d$ represent the depth of an attempted chain reorganization and $N$
represent the configured checkpoint depth. Under the checkpoint policy
adopted in this study, an attempted reorganization is permitted only when
its depth remains within the configured boundary. The decision rule is

\begin{equation}
C(d,N)=
\begin{cases}
\text{Accept}, & d \leq N,\\
\text{Reject}, & d > N.
\end{cases}
\label{eq:checkpoint-rule}
\end{equation}

Thus, reorganizations extending beyond the checkpoint boundary are
rejected, preventing the adversary from replacing deeper established
blockchain history under the considered model. The realized
reorganization-depth outcome can consequently be represented as

\begin{equation}
d_{\mathrm{realized}}=
\begin{cases}
d, & d \leq N,\\
0, & d > N,
\end{cases}
\label{eq:realized-depth}
\end{equation}

where a value of zero denotes a rejected reorganization attempt rather
than an actual zero-block reorganization.

The checkpoint depth controls the strictness of the mitigation mechanism.
A smaller $N$ establishes a stronger restriction on reorganizations,
whereas a larger $N$ permits greater reorganization flexibility. The
selection of $N$ therefore represents a trade-off between protection
against deep malicious reorganizations and tolerance for legitimate chain
reorganizations.

\subsection{Integrated Detection and Mitigation}
\label{subsec:integrated-defense}

The two components are integrated to address the two threat stages
identified in the threat model. During the first stage, the system
continuously observes hash-power concentration. When
$S_A(t)\geq\theta_w$, the state changes from normal to warning, providing
an indication that concentration is approaching the critical region. If
$S_A(t)\geq\theta_c$, the network is considered to have entered the
majority-control risk state.

The checkpoint mechanism provides the second defensive layer when an
adversary subsequently attempts a chain reorganization. The resulting
defensive process can therefore be expressed as

\begin{equation}
\begin{aligned}
S_A(t)
&\xrightarrow{\theta_w}
\text{Early Warning}
\xrightarrow{\theta_c}
\text{Majority-Control Risk}
\\
&\xrightarrow{N}
\text{Reorganization Control}.
\end{aligned}
\label{eq:integrated-defense}
\end{equation}

This integration is important because the two mechanisms address
different security objectives. Hash-power monitoring does not directly
prevent chain reorganization; instead, it provides advance indication of
increasing consensus concentration. Conversely, checkpointing does not
prevent an adversary from accumulating hash power; it limits the effect
of a subsequent attempt to modify established chain history. Combining
the two therefore provides a layered defense in which emerging risk is
identified before the critical condition and the potential impact of a
later reorganization is constrained.

The proposed approach does not assume that a single warning threshold or
checkpoint depth is universally optimal. Both $\theta_w$ and $N$ are
configurable parameters whose appropriate values depend on network
behavior and security requirements. The experimental evaluation therefore
examines multiple warning thresholds and checkpoint depths to determine
how these parameters influence early-warning capability and
reorganization resistance.

\section{Experiments}

The experiments evaluate the two components of the proposed defense:
(i) early detection of increasing hash-power concentration and
(ii) checkpoint-based mitigation of chain-reorganization attempts.
Experiment~1 evaluates the warning mechanism, while Experiment~2
evaluates the effect of checkpoint depth on reorganization resistance of 51\% attacks in proof-of-work blockchains.

\subsection{Experimental Setup}
\subsubsection*{Simulation Environment}
The experiments were conducted using Python, with custom scripts
developed for synthetic data generation, Monte Carlo simulation,
statistical analysis, and visualization. The simulation environment
was configured using predefined hash-power trajectories, stochastic
noise levels, warning and critical thresholds, checkpoint depths,
and reorganization-depth distributions. A fixed random seed was
used where applicable to support reproducibility.

\subsubsection*{Data}
The data used in these experiments was synthetically generated to enable controlled testing of the monitoring and mitigation strategies under repeatable conditions. Two distinct datasets were produced as is shown in Table~\ref{tab:datasets}.

The first dataset is a synthetic hash rate time series representing the distribution of mining power among multiple pools. In this dataset, Pool~A’s share of the network gradually increases from an initial 30\% to a peak of 58\%, with small stochastic noise added at each time step to emulate natural variations in mining power due to factors such as fluctuating miner participation and network latency. The shares of the remaining pools adjust dynamically to ensure that the total network hash rate remains constant at 100\%. This design allows for the evaluation of early-warning alert mechanisms as Pool~A approaches the 45\% (warning) and 50\% (critical) thresholds.

The second dataset models attack reorganization depths, which represent the number of blocks an adversary attempts to replace during a chain reorganization event. These values were sampled uniformly from the discrete range $\{3,4,\dots,13\}$ blocks, with repeated trials conducted to assess the variability and frequency of deep reorganizations. This dataset was used to compare outcomes under two scenarios: without checkpointing (where longer reorganizations may be accepted) and with checkpointing (where reorganizations deeper than a set finality limit $N=6$ are rejected. Table~\ref{tab:datasets} summarizes the main characteristics of the
synthetic datasets used in the experiments.

\begin{table}[!t]
\centering
\caption{Summary of Synthetic Datasets Used in Experiments}
\label{tab:datasets}
\footnotesize
\setlength{\tabcolsep}{3pt}
\renewcommand{\arraystretch}{1.05}

\begin{tabularx}{\columnwidth}{|p{0.30\columnwidth}|X|}
\hline
\textbf{Dataset} & \textbf{Description} \\ 
\hline

Hash-rate time series &
Simulated 60-minute series in which Pool~A increases from 30\% to
58\%, with stochastic noise. Used to evaluate the 45\% warning and
50\% critical thresholds. \\
\hline

Attack reorganization depths &
Simulated attack attempts with reorganization depths from 3 to
13 blocks. Used to evaluate checkpoint-based mitigation. \\
\hline

\end{tabularx}
\end{table}

\subsubsection*{Assumptions}
The experimental design was based on a set of controlled assumptions intended to isolate the variables of interest and ensure the repeatability of results. First, the blockchain difficulty adjustment mechanism was fixed for the duration of each experiment, meaning that no short-term difficulty retargeting effects were simulated. This choice eliminates variability in block production times that might otherwise occur due to changes in mining power, allowing for a more direct assessment of how hash rate concentration affects alert triggers and the feasibility of chain reorganizations. Second, transaction values within the simulated blocks were drawn from a uniform distribution, ensuring that the economic value of transactions did not influence miner behavior or attack strategies during any single run. By standardizing transaction values, the experiments focused solely on the structural and computational aspects of the network, rather than introducing additional complexity from profit-driven decision-making. These assumptions, while simplifying the experimental environment, make it possible to systematically analyze the technical effectiveness of the proposed monitoring and mitigation strategies without confounding factors related to fluctuating difficulty levels or economic incentives.



\subsection{Experiment 1: Hash-Power Monitoring and Early-Warning Detection}
\label{sec:exp1}

\subsubsection{Objective and Experimental Design}

Experiment~1 evaluates whether continuous monitoring of mining-pool
hash-power concentration can identify an emerging majority-control
condition before the critical threshold is reached. The primary metric
is the warning-to-critical intervention time, together with detection
behavior under different hash-power conditions.

The experiment models the hash-power share of a potentially dominant
mining pool, denoted as Pool~A, over a 60-minute simulation horizon.
Pool~A begins with approximately 30\% of the network hash power and
progressively increases its share toward 58\%. Short-term variation is
introduced using zero-mean stochastic noise with standard deviation
$\sigma=0.01$. The resulting hash-power share is represented as

\begin{equation}
s_A(t)=s_0+\frac{(s_f-s_0)t}{T-1}+\epsilon_t,
\label{eq:hash-power-model}
\end{equation}

where $s_0=0.30$, $s_f=0.58$, $T=60$, and $\epsilon_t$ represents the
stochastic component.

Two monitoring thresholds are defined. A warning is generated when

\begin{equation}
s_A(t)\geq 0.45,
\label{eq:warning-threshold}
\end{equation}

while the critical majority-control condition is represented by

\begin{equation}
s_A(t)\geq 0.50.
\label{eq:critical-threshold}
\end{equation}

Let $T_W$ and $T_C$ denote the first warning- and critical-threshold
crossing times, respectively. The available intervention lead time is

\begin{equation}
T_{\mathrm{lead}}=T_C-T_W.
\label{eq:lead-timer}
\end{equation}

The experiment was repeated over 1,000 Monte Carlo runs using a fixed
random seed for reproducibility. Each run generated an independent
realization of the stochastic component while preserving the same
underlying attack progression.

\begin{figure*}[!t]
    \centering
    \includegraphics[width=\textwidth]{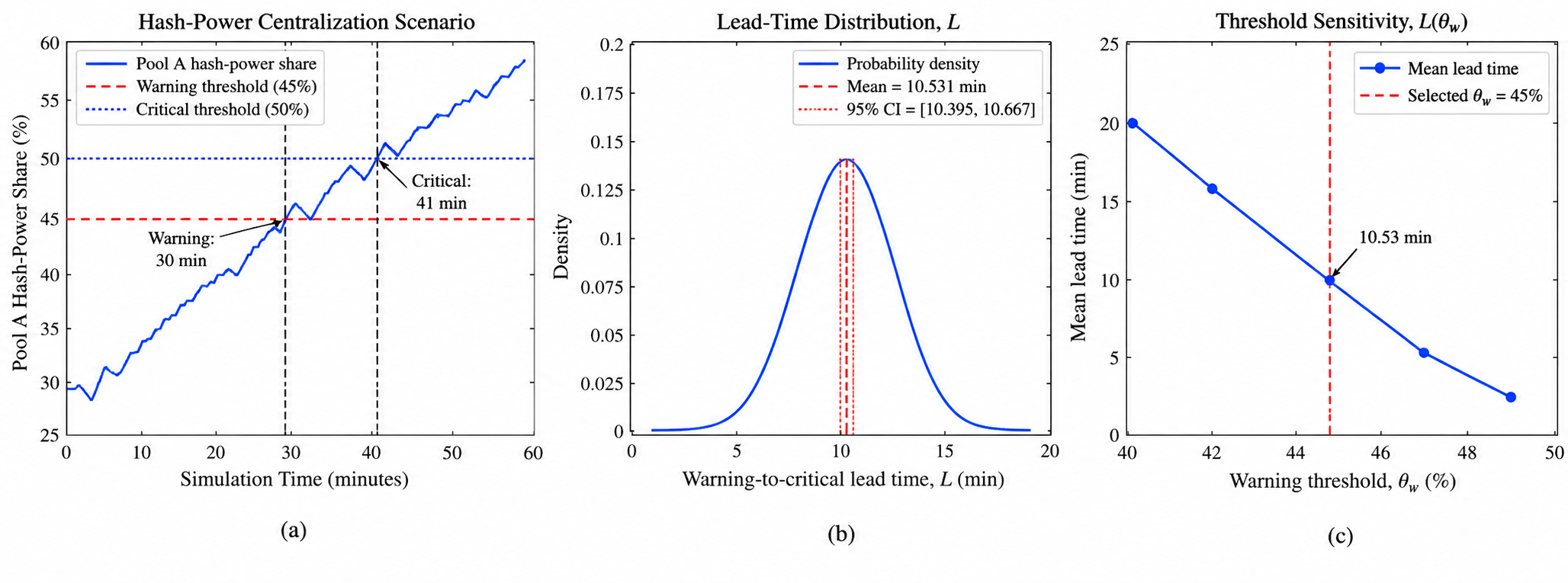}
    \caption{Evaluation of the early hash-power detection mechanism.
    (a) Representative hash-power centralization scenario showing the
    increase in Pool~A's share and the crossings of the 45\% warning and
    50\% critical thresholds.
    (b) Distribution of the warning-to-critical lead time, characterized
    by a mean of 10.531 minutes and a 95\% confidence interval of
    $[10.395,10.667]$ minutes.
    (c) Sensitivity of the mean lead time to the warning threshold
    $\theta_w$, with the selected 45\% operating point highlighted.}
    \label{fig:early-detection-results}
\end{figure*}

\begin{figure*}[!t]
    \centering
    \includegraphics[width=\textwidth]{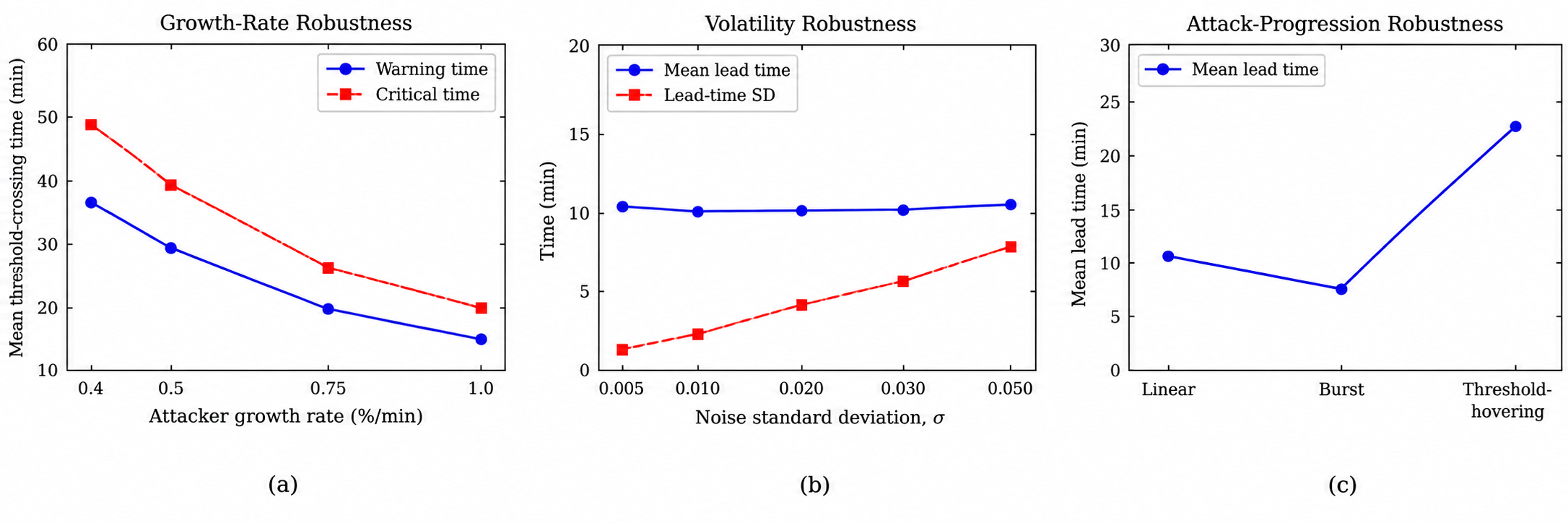}
    \caption{Robustness evaluation of the early hash-power detection
    mechanism.
    (a) Effect of attacker hash-power growth rate on warning and critical
    threshold-crossing times.
    (b) Effect of hash-power volatility on mean lead time and its
    variability.
    (c) Effect of attack-progression pattern on the warning-to-critical
    lead time.}
    \label{fig:exp1-robustness}
\end{figure*}

Figure~\ref{fig:early-detection-results}(a) illustrates a representative attack trajectory. The warning threshold is first crossed at approximately minute~30 and the critical threshold at approximately minute~41, producing an 11-minute warning interval. This trajectory is illustrative; the statistical evaluation is based on all 1,000 Monte Carlo runs.

\subsubsection{Monte Carlo Results}

Across the 1,000 simulated attack trajectories, the mean warning time was 30.982 minutes and the mean critical time was 41.513 minutes. The resulting mean warning-to-critical lead time was 10.531 minutes, with a median of 11.000 minutes and a standard deviation of 2.202 minutes as is shown in Table \ref{Montes}. Monte Carlo simulation was selected to evaluate the proposed mechanisms across repeated stochastic conditions, reducing reliance on individual simulation outcomes and enabling robust statistical analysis \cite{mooney1997monte}.

The 95\% confidence interval for the mean intervention time was
$[10.395,10.667]$ minutes. Figure~\ref{fig:early-detection-results} summarizes the early-detection results.
Figure~\ref{fig:early-detection-results}(a) illustrates the warning and critical
threshold crossings, while Fig.~\ref{fig:early-detection-results}(b) shows the
resulting lead-time distribution with most observations
concentrated around the 10--12 minute region. Fig.~\ref{fig:early-detection-results}(c)
shows how the warning threshold influences the available intervention time.


\begin{table}[!t]
\centering
\caption{Summary of Experiment~1 Monte Carlo results.}
\label{tab:exp1-results}
\begin{tabular}{lr}
\hline
\textbf{Metric} & \textbf{Result} \\
\hline
Monte Carlo runs & 1,000 \\
Mean warning time & 30.982 min \\
Mean critical time & 41.513 min \\
Mean lead time & 10.531 min \\
Median lead time & 11.000 min \\
Lead-time standard deviation & 2.202 min \\
95\% CI of mean lead time & [10.395, 10.667] min \\
\hline
\label{Montes}
\end{tabular}
\end{table}

\subsubsection{Robustness Analysis}
\label{subsubsec:exp1-robustness}

Robustness was evaluated under variations in attacker hash-power growth rate, hash-power volatility, and attack-progression pattern.Table~\ref{tab:exp1-robustness} summarizes the robustness results under
different hash-power growth rates, volatility levels, and attack-progression patterns.

Figure~\ref{fig:exp1-robustness}(a) shows that faster hash-power
accumulation reduces the available intervention time. The mean lead time
decreases from 12.530 minutes at 0.40\%/min to 4.970 minutes at
1.00\%/min. Warning and critical detection nevertheless remain at
100\% across all evaluated growth rates.

Figure~\ref{fig:exp1-robustness}(b) shows that increasing volatility has
little effect on the expected lead time, which remains approximately
10.5 minutes, but substantially increases its variability. The standard
deviation rises from 1.294 minutes at $\sigma=0.005$ to 7.716 minutes
at $\sigma=0.050$. Thus, volatile hash-power conditions make the
available response interval less predictable.

Figure~\ref{fig:exp1-robustness}(c) evaluates different attack-progression
patterns. The linear and burst scenarios produce mean lead times of
10.583 and 7.490 minutes, respectively. The threshold-hovering scenario
produces a longer mean lead time of 22.682 minutes because the simulated
attacker remains close to the warning region for an extended period
before reaching the critical threshold.

\begin{table*}[!t]
\centering
\caption{Robustness analysis of the early hash-power detection mechanism.}
\label{tab:exp1-robustness}
\begin{tabular}{llccc}
\hline
\textbf{Analysis} &
\textbf{Configuration} &
\textbf{Critical Detection (\%)} &
\textbf{Mean Lead Time (min)} &
\textbf{SD (min)} \\
\hline
Growth Rate
& 0.40\%/min & 100 & 12.530 & 2.618 \\
& 0.50\%/min & 100 & 9.964 & 2.163 \\
& 0.75\%/min & 100 & 6.648 & 1.536 \\
& 1.00\%/min & 100 & 4.970 & 1.234 \\
\hline
Volatility
& $\sigma=0.005$ & 100 & 10.565 & 1.294 \\
& $\sigma=0.010$ & 100 & 10.441 & 2.230 \\
& $\sigma=0.020$ & 100 & 10.441 & 3.972 \\
& $\sigma=0.030$ & 100 & 10.474 & 5.496 \\
& $\sigma=0.050$ & 100 & 10.638 & 7.716 \\
\hline
Attack Progression
& Linear & 100 & 10.583 & 2.115 \\
& Burst & 100 & 7.490 & 1.683 \\
& Threshold-hovering & 100 & 22.682 & 5.800 \\
\hline
\end{tabular}
\end{table*}

The mechanism detects the modeled critical progression across all
evaluated robustness conditions, but the available intervention window is
strongly influenced by the temporal behavior of hash-power concentration.

\subsubsection{Joint Sensitivity to Hash-Power Growth and Volatility}
\label{subsubsec:joint-sensitivity}

To further evaluate the robustness of the early-warning mechanism,
a joint sensitivity experiment was conducted in which hash-power
growth rate and stochastic volatility were varied simultaneously.
Five growth rates,
$\{0.40,0.50,0.60,0.75,1.00\}\%/\mathrm{min}$, were evaluated
across five noise levels,
$\sigma\in\{0.005,0.010,0.020,0.030,0.050\}$.
Each parameter combination was evaluated using 1,000 Monte Carlo
runs while retaining the 45\% warning and 50\% critical thresholds.

\begin{figure*}[!t]
    \centering
    \includegraphics[width=\textwidth]{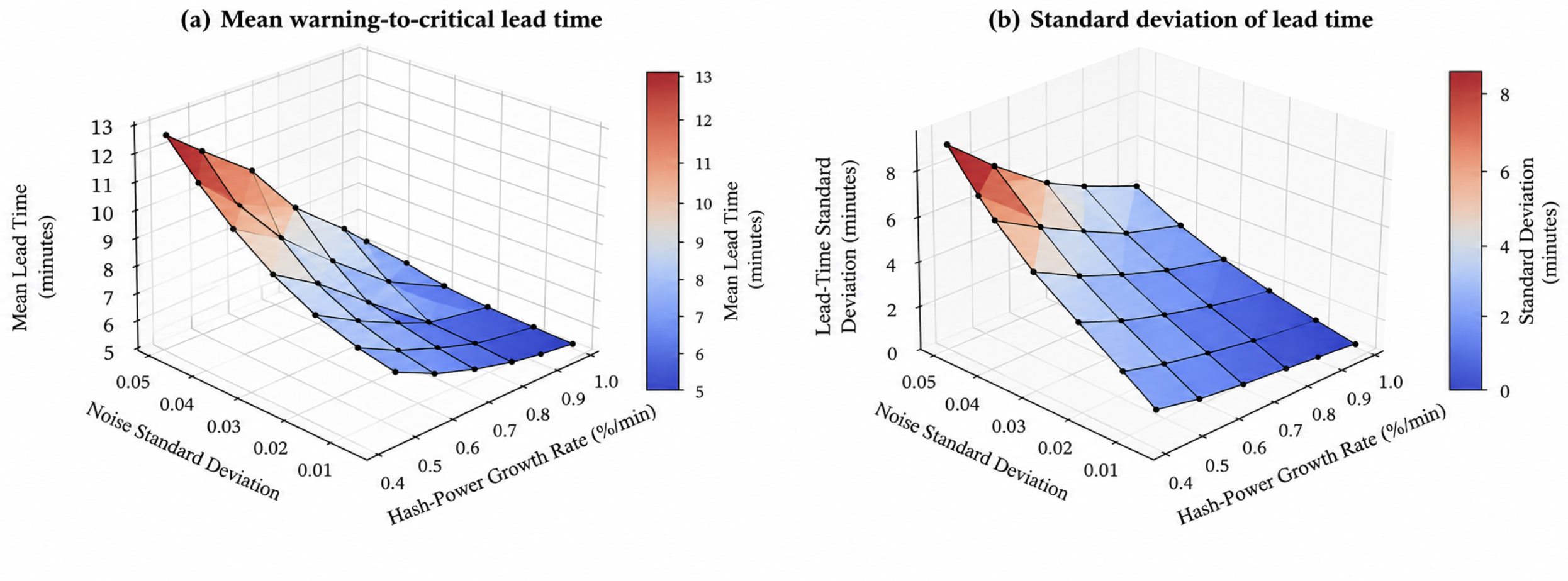}
    \caption{Joint sensitivity of the early-warning mechanism to
    hash-power growth rate and stochastic volatility.
    (a) Mean warning-to-critical lead time.
    (b) Standard deviation of the warning-to-critical lead time.
    Each parameter combination was evaluated using 1,000 Monte Carlo
    runs.}
    \label{fig:joint-sensitivity}
\end{figure*}

Fig.~\ref{fig:joint-sensitivity} presents the joint sensitivity
results. As shown in Fig.~\ref{fig:joint-sensitivity}(a), the mean
warning-to-critical lead time decreases primarily as the hash-power
growth rate increases. At a growth rate of
$0.40\%/\mathrm{min}$, the mean lead time remains approximately
12.4--12.6 minutes across the evaluated volatility levels, whereas
at $1.00\%/\mathrm{min}$ it decreases to approximately
4.9--5.0 minutes. Increasing volatility, in contrast, produces
comparatively small changes in the mean lead time.

The effect of volatility becomes more apparent in
Fig.~\ref{fig:joint-sensitivity}(b). Increasing the noise level
substantially increases the standard deviation of the
warning-to-critical lead time, particularly at slower hash-power
growth rates. For example, at $0.40\%/\mathrm{min}$, the standard
deviation increases from approximately 1.45 minutes at
$\sigma=0.005$ to approximately 9.03 minutes at
$\sigma=0.050$. Thus, although volatility has limited influence on
the expected lead time, it substantially reduces the predictability
of the available intervention window.

It has been observed that, the joint analysis indicates that attack progression rate
primarily determines the expected intervention window, whereas
hash-power volatility primarily determines the variability of that
window. At the most volatile configuration
($0.40\%/\mathrm{min}$, $\sigma=0.050$), 999 of the 1,000
simulations produced a valid warning-to-critical sequence, further
illustrating the effect of high volatility on detection consistency.

\subsubsection{Behavior Under Different Network Conditions}

To examine behavior beyond a continuously increasing attack trajectory,
three network conditions were considered: \emph{normal}, \emph{recovery},
and \emph{attack}. The normal scenario increases from approximately
30\% to 38\%, remaining below the warning threshold. The recovery
scenario temporarily reaches approximately 47\% before declining toward
35\%, while the attack scenario progresses from approximately 30\% to
58\%.

Each condition was evaluated over 1,000 runs. All attack trajectories
generated both warning and critical conditions, while none of the normal
trajectories triggered either condition. Recovery trajectories generated
warnings because their hash-power share genuinely entered the warning
region, but only 0.3\% subsequently crossed the critical threshold due to
stochastic variation. These events are therefore treated as
\emph{non-escalating warnings} rather than conventional false positives.

\subsubsection{Warning-Threshold Sensitivity}

Five warning thresholds were evaluated: 40\%, 42.5\%, 45\%, 47.5\%,
and 49\%. Table~\ref{tab:threshold-analysis} summarizes their detection
behavior and corresponding intervention times.

\begin{table}[!t]
\centering
\caption{Sensitivity of the monitoring mechanism to the warning threshold.}
\label{tab:threshold-analysis}
\resizebox{\linewidth}{!}{
\begin{tabular}{ccccc}
\hline
\textbf{Threshold} &
\textbf{Attack Detection} &
\textbf{Normal False Warning} &
\textbf{Recovery Warning} &
\textbf{Mean Lead Time} \\
\hline
40.0\% & 100\% & 7.5\% & 100\% & 21.06 min \\
42.5\% & 100\% & 0\% & 100\% & 15.67 min \\
45.0\% & 100\% & 0\% & 100\% & 10.53 min \\
47.5\% & 100\% & 0\% & 71.7\% & 5.26 min \\
49.0\% & 100\% & 0\% & 6.4\% & 2.09 min \\
\hline
\end{tabular}}
\end{table}

Figure~\ref{fig:early-detection-results}(c) illustrates the principal
trade-off. Lower warning thresholds provide greater intervention time,
whereas thresholds closer to 50\% substantially shorten the response
window. For example, increasing the warning threshold from 40\% to
49\% reduces the mean lead time from 21.06 to 2.09 minutes.

Table~\ref{tab:threshold-analysis} also shows the behavior under normal
and recovery conditions. The 40\% threshold produces warnings in 7.5\%
of normal simulations, while thresholds of 42.5\% and above eliminate
normal warnings under the modeled conditions. At 49\%, however, only
6.4\% of recovery trajectories are detected.

Under the evaluated synthetic conditions, the 45\% threshold provides a
useful experimental operating point: it detects all modeled attack
trajectories, produces no warnings during the normal scenario, and retains
a mean intervention interval of approximately 10.53 minutes. This result
does not establish 45\% as a universally optimal threshold.

\subsection{Experiment 2: Checkpoint-Based Reorganization Resistance}
\label{sec:exp2}

\subsubsection{Objective and Experimental Configuration}

Experiment~2 evaluates checkpoint-based finality as a mechanism for
limiting chain-reorganization attempts. The experiment examines how the
checkpoint depth $N$ influences both attack rejection and the resulting
reorganization-depth outcome.

A total of 1,000 attempted reorganizations were generated with depths

\begin{equation}
L \in \{3,4,\ldots,13\},
\label{eq:attack-depths}
\end{equation}

where $L$ denotes the attempted reorganization depth in blocks. The same
attack-generation procedure was used for each checkpoint configuration to
enable direct comparison.

Five baseline checkpoint depths were evaluated,

\begin{equation}
N \in \{2,4,6,8,10\},
\label{eq:checkpoint-depths}
\end{equation}

together with a no-checkpoint baseline. An attempted reorganization is
accepted when

\begin{equation}
L\leq N,
\end{equation}

and rejected when

\begin{equation}
L>N.
\end{equation}

Rejected attempts are represented by a reorganization-depth outcome of
zero. Therefore, a zero value denotes a rejected attack rather than an
actual zero-block reorganization.

Repeated Monte Carlo batches were used to evaluate the stability of the
results. Each checkpoint configuration was evaluated across 1,000
observations.

\begin{figure*}[!t]
    \centering
    \includegraphics[width=\textwidth]{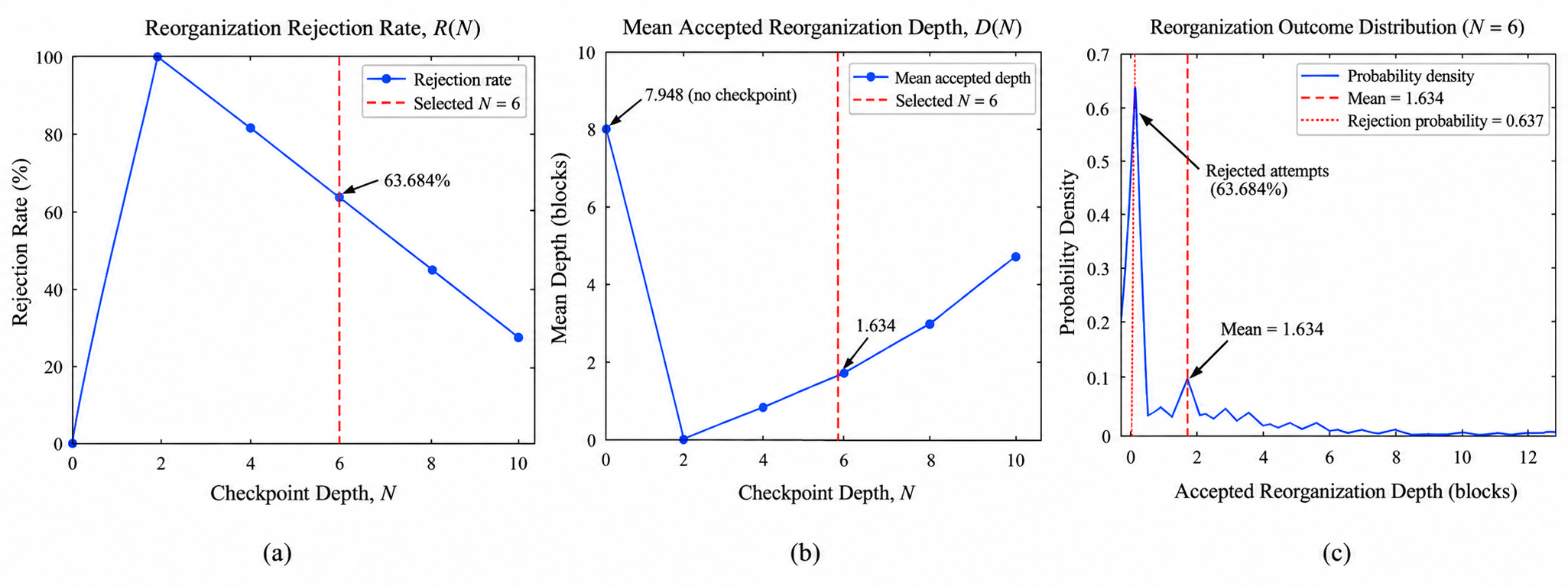}
    \caption{Evaluation of the checkpoint-based mitigation mechanism.
    (a) Reorganization rejection rate as a function of checkpoint depth
    $N$.
    (b) Mean reorganization-depth outcome under different checkpoint
    depths compared with the no-checkpoint baseline.
    (c) Distribution of reorganization outcomes for $N=6$, where zero
    represents a rejected attack attempt.}
    \label{fig:exp2-checkpoint-results}
\end{figure*}

\begin{figure*}[!t]
    \centering
    \includegraphics[width=\textwidth]{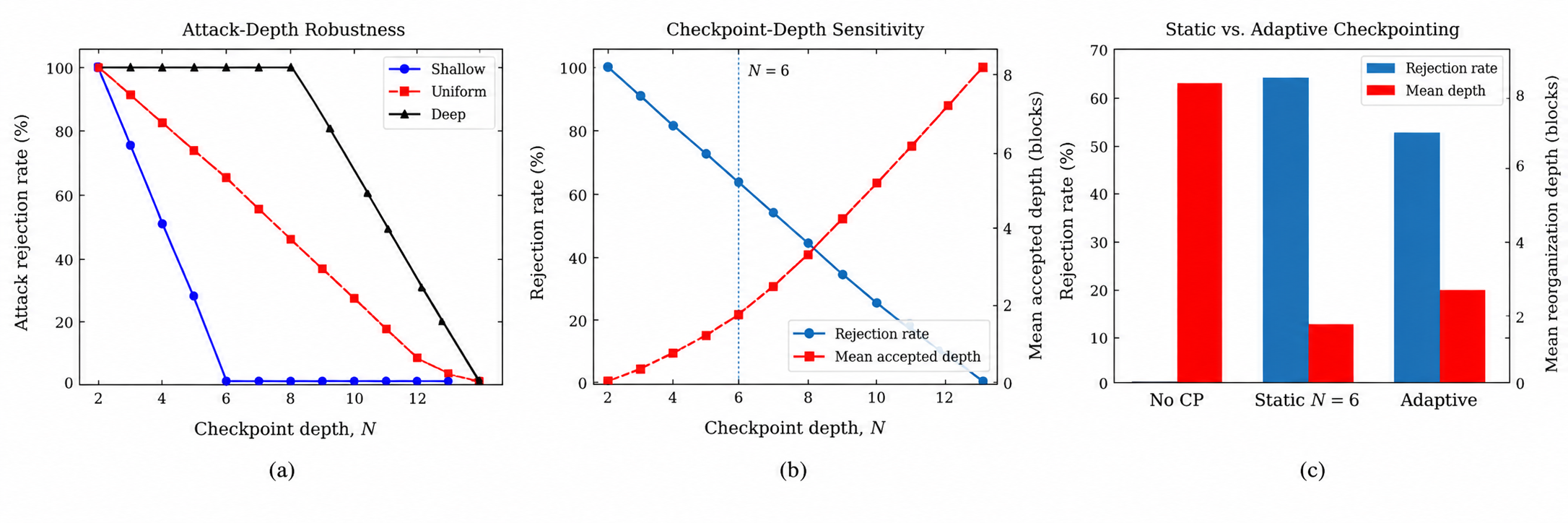}
    \caption{Robustness and sensitivity evaluation of checkpoint-based
    mitigation.
    (a) Effect of attack-depth distribution on rejection rate.
    (b) Relationship between checkpoint depth, attack rejection, and
    reorganization-depth outcome.
    (c) Comparison of no checkpointing, static checkpointing at $N=6$,
    and the evaluated adaptive checkpointing policy.}
    \label{fig:exp2-robustness}
\end{figure*}

\subsubsection{Checkpointing Results}

Table~\ref{tab:checkpoint-results} summarizes the baseline checkpoint
results.

\begin{table}[!t]
\centering
\caption{Monte Carlo results for different checkpoint depths.}
\label{tab:checkpoint-results}
\resizebox{\columnwidth}{!}{
\begin{tabular}{c|c|c|c}
\hline
\textbf{Checkpoint} &
\textbf{Acceptance (\%)} &
\textbf{Rejection (\%)} &
\textbf{Mean Depth} \\
\hline
$N=2$  & 0.000  & 100.000 & 0.000 \\
$N=4$  & 18.160 & 81.840  & 0.636 \\
$N=6$  & 36.316 & 63.684  & 1.634 \\
$N=8$  & 54.519 & 45.481  & 3.000 \\
$N=10$ & 72.701 & 27.299  & 4.727 \\
\hline
\end{tabular}}
\end{table}

As shown in Fig.~\ref{fig:exp2-checkpoint-results}(a), rejection decreases
monotonically as $N$ increases. The $N=2$ configuration rejects all
simulated attempts because the minimum generated attack depth is three
blocks. At $N=6$, the rejection rate is 63.684\%, while at $N=10$ it
decreases to 27.299\%.

Figure~\ref{fig:exp2-checkpoint-results}(b) shows the corresponding
reorganization-depth outcomes. The no-checkpoint baseline produces a mean
depth of 7.948 blocks. At $N=6$, this decreases to 1.634 blocks,
representing an approximately 79.4\% reduction under the adopted
measurement convention.

Because rejected attacks are encoded as zero, the reported mean represents
the aggregate outcome over all attempts rather than the conditional depth
of successful reorganizations.

\subsubsection{Robustness and Checkpoint Sensitivity}

Figure~\ref{fig:exp2-robustness}(a) evaluates different attack-depth
distributions. The results show that checkpoint effectiveness depends not
only on $N$ but also on the depth distribution of attempted
reorganizations. Shallow attacks become progressively admissible as $N$
increases, whereas deep attacks remain rejected under stricter checkpoint
settings.

The full checkpoint-depth sweep in
Fig.~\ref{fig:exp2-robustness}(b) confirms the security--flexibility
trade-off. Increasing $N$ reduces the rejection rate while increasing the
mean reorganization-depth outcome. At the experimental operating point
$N=6$, the rejection rate is approximately 63.60\% and the mean outcome
is approximately 1.638 blocks.

Figure~\ref{fig:exp2-robustness}(c) compares the no-checkpoint baseline,
static checkpointing at $N=6$, and the evaluated adaptive policy. Static
checkpointing achieves a rejection rate of 63.60\% and mean depth of
1.637 blocks, whereas the adaptive configuration achieves 53.18\% and
2.652 blocks, respectively. Under the evaluated conditions, static
checkpointing therefore provides stronger reorganization resistance.
The adaptive configuration should instead be interpreted as an
illustrative risk-responsive policy rather than an optimized strategy.

\begin{table*}[!t]
\centering
\caption{Summary of checkpoint robustness and sensitivity analysis.}
\label{tab:exp2-robustness}
\begin{tabular}{llccc}
\hline
\textbf{Analysis} &
\textbf{Configuration} &
\textbf{Rejection (\%)} &
\textbf{Mean Depth (blocks)} &
\textbf{Mitigation Gain (\%)} \\
\hline
Checkpoint Sweep
& $N=2$  & 100.00 & 0.000 & 100.00 \\
& $N=4$  & 81.80  & 0.637 & 92.04 \\
& $N=6$  & 63.60  & 1.638 & 79.52 \\
& $N=8$  & 45.48  & 2.996 & 62.54 \\
& $N=10$ & 27.28  & 4.730 & 40.88 \\
& $N=12$ & 9.08   & 6.817 & 14.79 \\
& $N=13$ & 0.00   & 7.996 & 0.05 \\
\hline
Configuration
& No checkpoint & 0.00  & 7.995 & -- \\
& Static $N=6$  & 63.60 & 1.637 & -- \\
& Adaptive      & 53.18 & 2.652 & -- \\
\hline
\end{tabular}
\end{table*}

\subsubsection{Security--Flexibility Trade-off}

The results demonstrate that checkpoint depth is a configurable security
parameter rather than a universally optimal constant. Smaller values of
$N$ provide stronger resistance to deep reorganization attempts, whereas
larger values permit greater tolerance for legitimate chain changes.

The $N=6$ configuration is therefore treated as an intermediate
experimental operating point rather than an optimal value. Combined with
Experiment~1, the results support a layered defense in which increasing
hash-power concentration is detected before the critical region is reached,
while checkpointing constrains the impact of subsequent chain-reorganization
attempts.



\section{Results and Discussion}
\label{sec:results-discussion}

The experimental results provide evidence for two complementary security
functions of the proposed approach. First, monitoring hash-power
concentration can provide advance indication of an emerging
majority-control condition. Second, checkpoint-based finality can constrain
the impact of subsequent chain-reorganization attempts. The results also
show that the effectiveness of both mechanisms depends strongly on their
parameter settings and on the dynamics of the modeled attack.

\subsection{Effectiveness of Early Hash-Power Detection}
\label{subsec:early-detection}

Experiment~1 demonstrates that monitoring hash-power concentration below
the critical majority-control boundary can provide a measurable
intervention window. At the selected 45\% warning threshold, the mean
warning-to-critical lead time was 10.531 minutes across 1,000 Monte Carlo
runs, with a 95\% confidence interval of $[10.395,10.667]$ minutes.
Rather than indicating that an attack has succeeded, the warning therefore
serves as an indicator that mining concentration has entered a region that
may justify closer monitoring or defensive preparation.

The robustness analysis further shows that the usefulness of this warning
depends on how hash-power concentration develops. Faster accumulation
reduced the mean intervention window from 12.530 minutes at 0.40\%/min
to 4.970 minutes at 1.00\%/min. Increased volatility had comparatively
little effect on the mean lead time but substantially increased its
variability. Burst progression also shortened the available response
interval, whereas threshold-hovering produced a longer period of elevated
concentration before the critical boundary was reached. These findings
indicate that detection performance should be evaluated not only by whether
a threshold is crossed, but also by the amount and predictability of the
response time that precedes critical concentration.

The network-condition experiments provide an additional interpretation.
Warnings generated during recovery scenarios should not automatically be
classified as erroneous detections because the modeled hash-power share
genuinely entered the warning region. Such events are better interpreted
as non-escalating warnings: elevated concentration was observed, but the
condition subsequently returned to a lower-risk state. This distinction is
important because the warning mechanism identifies elevated concentration
risk rather than predicting with certainty that majority control will
subsequently occur.

\subsection{Effectiveness of Checkpoint-Based Mitigation}
\label{subsec:checkpoint-effectiveness}

Experiment~2 demonstrates that checkpointing can substantially constrain
the outcome of simulated chain-reorganization attempts. Without
checkpointing, the mean reorganization depth was 7.948 blocks. Introducing
a checkpoint boundary reduced both the proportion of admissible
reorganizations and their aggregate depth.

The results reveal a monotonic relationship between checkpoint depth and
reorganization resistance. More restrictive checkpoint settings produced
higher rejection rates and smaller reorganization-depth outcomes. At the
experimental operating point $N=6$, approximately 63.684\% of the
simulated attempts were rejected and the mean reorganization-depth outcome
was reduced to 1.634 blocks, corresponding to an approximately 79.4\%
reduction relative to the no-checkpoint baseline.

This reduction should be interpreted together with the adopted measurement
convention. Rejected attacks are represented by a depth of zero; therefore,
the reported mean captures the combined effect of attack rejection and the
depth of attempts that remain admissible. It is not the conditional mean
depth of successful reorganizations alone.

The robustness analysis further demonstrates that checkpoint effectiveness
depends on the characteristics of the attempted attack. A checkpoint depth
that strongly rejects deeper reorganizations may permit a larger proportion
of shallow attempts. Consequently, checkpoint depth cannot be evaluated
independently of the expected attack-depth distribution. The comparison
between static and adaptive checkpointing reinforces this observation.
Under the evaluated conditions, static checkpointing at $N=6$ provided
stronger reorganization resistance than the illustrative adaptive policy.
The adaptive result should therefore be interpreted as evidence of the
feasibility of risk-responsive checkpoint selection rather than evidence
that adaptation inherently improves security.

The joint sensitivity analysis further confirms this distinction.
Across simultaneous variations in growth rate and volatility,
hash-power growth rate primarily determined the expected intervention
window, whereas increasing volatility predominantly increased the
dispersion of lead times. This suggests that rapidly increasing
concentration reduces the time available for intervention, while
volatile concentration makes that available response time less
predictable.

\subsection{Combined Security Implications}
\label{subsec:combined-security}

The two experiments address different stages of majority-control risk.
Experiment~1 addresses the development and detection of excessive
hash-power concentration, whereas Experiment~2 addresses the consequences
of a subsequent chain-reorganization attempt. Their relationship can be
summarized as

\begin{equation}
\begin{aligned}
\text{Monitoring}
&\rightarrow \text{Early Warning}
\rightarrow \text{Critical Detection}
\\
&\rightarrow \text{Checkpointing}
\rightarrow \text{Reorganization Control}.
\end{aligned}
\label{eq:combined-defense}
\end{equation}

The two mechanisms should not be interpreted as substitutes. Hash-power
monitoring does not itself prevent chain reorganization, and checkpointing
does not prevent mining-power concentration. Instead, monitoring provides
information about the development of a potentially dangerous condition,
while checkpointing constrains the extent to which chain history can be
modified under the adopted finality policy.

Under the selected experimental operating points, the 45\% warning
threshold provided approximately 10.53 minutes of mean advance warning,
while checkpointing at $N=6$ rejected approximately 63.68\% of the modeled
reorganization attempts and reduced the mean reorganization-depth outcome
to approximately 1.63 blocks. These measurements represent distinct
security properties and should not be combined into a single performance
metric. Collectively, however, they support the layered design principle of
combining early risk indication with impact limitation.

It is important to distinguish this layered interpretation from a fully
coupled defense evaluation. The two mechanisms were experimentally
evaluated as separate components of the proposed framework. The results
therefore demonstrate complementary detection and mitigation behavior,
rather than proving that the combined system prevents 51\% attacks under
all network conditions.

\subsection{Sensitivity and Security Trade-offs}
\label{subsec:tradeoffs}

Both experiments reveal parameter-dependent security trade-offs. In
Experiment~1, lowering the warning threshold increases the available
intervention time but may also increase unnecessary warnings. The 40\%
threshold, for example, provided a mean lead time of 21.06 minutes but
generated warnings in 7.5\% of the modeled normal scenarios. Moving the
threshold closer to 50\% reduced such sensitivity but sharply shortened
the intervention window, which fell to only 2.09 minutes at 49\%.

A corresponding trade-off occurs in checkpoint selection. Smaller values
of $N$ provide stronger finality and reject a larger proportion of the
modeled reorganizations, whereas larger values permit greater
reorganization flexibility. The extreme $N=2$ result, for example, should
not be interpreted as universal protection: all attacks were rejected
because the simulated attack depths began at three blocks. This illustrates
the dependence of checkpoint effectiveness on both parameter selection and
the evaluated attack model.

The warning threshold and checkpoint depth should therefore be treated as
configurable security parameters rather than universal constants. The
45\% warning threshold and $N=6$ checkpoint depth serve as experimental
operating points in this study. Their suitability in operational blockchain
networks would depend on mining behavior, acceptable alert frequency,
reorganization tolerance, network dynamics, and the required response time.

\subsection{Limitations and Threats to Validity}
\label{subsec:limitations}

The results in this work should be interpreted within the assumptions of the simulation
environment. The conducted experiments rely on synthetic hash-power trajectories and simulated reorganization depths rather than measurements from a live
blockchain network. Factors such as mining pool migration, network latency,
competing forks, miner incentives, and measurement uncertainty are therefore
only partially represented.

Experiment~1 used a predefined hash-power progression patterns and treats the 50\% level as the critical majority control boundary. Hash power is thus
used as a security indicator rather than as a complete predictor of
successful attack behavior. Experiment~2 restricts attempted reorganization depths to 3--13 blocks and represents rejected attempts by a
reorganization-depth outcome of zero.

The evaluation also does not quantify operational and economic costs,
including false-warning response, checkpoint-management overhead,
legitimate fork recovery, or miner incentives. Accordingly, these findings
demonstrate the behavior and trade-offs of the proposed mechanisms under
the evaluated conditions rather than universal performance guarantees.
Further evaluation using the network-level simulation and empirical blockchain
data is required to assess their effectiveness under more realistic
conditions.

\section{Conclusion and Future Work}
\label{sec:conclusion}

This study has investigated a two-layer defense against 51\% attacks in
Proof-of-Work blockchains by combining early hash-power monitoring with
checkpoint-based mitigation. The early-detection mechanism from experiment 1 demonstrated
that increasing hash-power concentration can provide an intervention
window before the critical majority-control threshold is reached. Across the conducted
1,000 Monte Carlo simulations, the selected 45\% warning threshold
provided a mean warning-to-critical lead time of 10.531 minutes.

The checkpoint-based mitigation further demonstrated the ability to constrain
chain-reorganization attempts. At the experimental operating point
$N=6$, approximately 63.684\% of simulated reorganization attempts were
rejected, while the mean reorganization-depth outcome was reduced from
7.948 blocks without checkpointing to 1.634 blocks. The sensitivity
analyses also showed that neither the warning threshold nor checkpoint
depth should be considered universally optimal; both represent
security flexibility trade-offs that depend on network conditions and
security requirements.

For future work, we aim to  evaluate the proposed mechanisms using network-level and empirical blockchain data, including historical mining-pool behavior and chain-reorganization events. Further research will also investigate adaptive warning thresholds and checkpoint policies that respond dynamically to changes in network risk, together with the operational and economic costs associated with stronger finality mechanisms.

\ifCLASSOPTIONcompsoc
  \section*{Acknowledgments}
\else
  \section*{Acknowledgment}
\fi

The author acknowledges the support of the Department of Computer
Science and Engineering at the University of Colorado Denver, USA.
The opinions, findings, and conclusions expressed in this article are
solely those of the author.

\ifCLASSOPTIONcaptionsoff
  \newpage
\fi

\IEEEtriggeratref{18}
\bibliographystyle{IEEEtran}
\bibliography{references}

\vspace{-10cm}

\begin{IEEEbiography}
[{\includegraphics[width=1in,height=1.25in,clip,keepaspectratio]{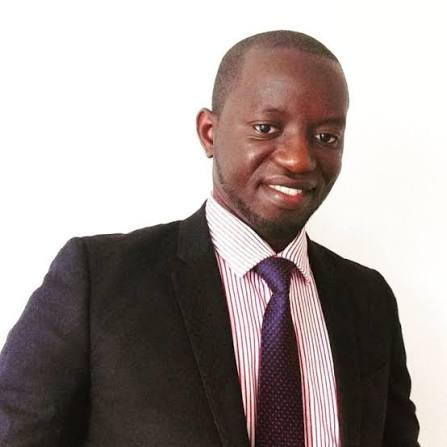}}]
{VICTOR KEBANDE}

is a cybersecurity researcher and Assistant Professor of Cybersecurity
at the University of Colorado Denver, Denver, CO, USA, and is also
affiliated with the ATLAS Institute, University of Colorado Boulder,
Boulder, CO, USA. His research interests include cybersecurity,
digital forensics in the Internet of Things, artificial intelligence
in cybersecurity, critical infrastructure protection, and cloud security.
Kebande received the Ph.D. degree in computer science, with a focus on
information and computer security architectures and digital forensics,
in 2018. He serves on the editorial board of \textit{Forensic Science
International: Reports}. He is a Member of IEEE. Contact him at
victor.kebande@ucdenver.edu or victor.kebande@colorado.edu.
\end{IEEEbiography}

\vspace{1em}

\noindent
\raisebox{0.1ex}{\rule{1.2ex}{1.2ex}}\hspace{0.5em}
Direct questions and comments about this article to Victor Kebande,
University of Colorado Denver, Denver, CO, USA;
\href{mailto:victor.kebande@ucdenver.edu}{victor.kebande@ucdenver.edu};
\href{mailto:victor.kebande@colorado.edu}{victor.kebande@colorado.edu}.

\end{document}